\documentclass[twocolumn,showpacs,eqsecnum,pra]{revtex4}

\usepackage{amsmath}

\usepackage{amssymb}
\usepackage{epsfig}

\begin{document}

\title {Gaussification through decoherence}
\author{Paulina Marian$^{1,2}$}
\email{paulina.marian@g.unibuc.ro}
\author{Iulia Ghiu{$^1$}}
\email{iulia.ghiu@g.unibuc.ro}
\author{ Tudor A. Marian$^{1}$}
\email{tudor.marian@g.unibuc.ro}
\affiliation{ $^1$Centre for Advanced  Quantum Physics,
Department of Physics, University of Bucharest, 
R-077125 Bucharest-M\u{a}gurele, Romania\\
$^2$Department of Physical Chemistry, University of Bucharest,\\
Boulevard Regina Elisabeta 4-12, R-030018  Bucharest, Romania}

\begin{abstract}

We investigate the loss of nonclassicality and  non-Gaussianity 
of a single-mode state of the radiation field in contact with a thermal reservoir.  The damped density matrix for a Fock-diagonal input is written using the Weyl expansion of the density operator. Analysis of the evolution of the quasiprobability densities reveals the existence of two successive characteristic times 
of the reservoir which are sufficient to assure the positivity of the Wigner function and, respectively, of the  $P$ representation. We examine the time evolution  of non-Gaussianity using three recently introduced distance-type measures. They are based on the Hilbert-Schmidt metric, the relative entropy, and the Bures metric.   Specifically, for an $M$-photon-added thermal state, 
we obtain a compact analytic formula of the time-dependent density matrix that is used to evaluate and compare the three non-Gaussianity measures. We find a good consistency of these measures on the sets of damped states. The explicit damped quasiprobability densities are shown to support our general findings regarding the loss of negativities of Wigner and $P$ functions during decoherence.  Finally, we point out that Gaussification 
of the attenuated field mode is accompanied by a nonmonotonic evolution of the von Neumann entropy of its state conditioned 
by the initial value of the mean photon number.

\end{abstract}


\maketitle

\section{Introduction}
In quantum optics, a nonclassical state is defined as having either a negative or a highly singular $P$ representation (more singular than  Dirac's $\delta$). Otherwise, we term it as a classical state. Cahill \cite{Cah} and later on Hillery \cite{Hill} proved that the only pure states that are classical are the coherent ones: all other classical states are mixtures. Accordingly, a pure state whose Wigner function has also negative values is neither classical nor Gaussian. On the other hand, Hudson's theorem \cite{Hud} implies that the  Wigner function of a non-Gaussian pure state 
is not pointwise non-negative. Thus, for pure states, negativity 
of the Wigner function could be interpreted as an indicator 
of nonclassicality as prominent as the negativity of the $P$ representation. Moreover, signatures of nonclassicality were identified by means of the negativity of the Wigner function 
for some mixed non-Gaussian states as well \cite{KZ}. Interest 
in the non-Gaussian states  has interestingly emerged in quantum information processing. Very recently it was shown 
that continuous-variable states and operations represented 
by non-negative (Gaussian or non-Gaussian) Wigner functions can be efficiently simulated on a classical computer \cite{ME,VWFE}. 
This result is a continuous-variable generalization of the Gottesman-Knill theorem for qubits \cite{NC}. At the same time, 
the negativity of the discrete Wigner function defined 
for odd-dimensional discrete-variable systems was identified to be a quantum computational resource \cite{VFGE,ME}. This means that the states with negative Wigner functions cannot be efficiently simulated on a classical computer or, otherwise said, a quantum computer outperforms a classical one. This property was extended 
to the infinite-dimensional case  \cite{ME} and examples of useful resource states for quantum computation were given \cite{VWFE}.
Some time ago, it was also realized that non-Gaussian resources 
and operations could be more efficient in some quantum  protocols such as teleportation \cite{Opat,Paris1,Il} and cloning 
\cite{Cerf1} or even indispensable in entanglement distillation \cite{Plenio}. To quantify the non-Gaussianity as a resource in such cases, some distance-type measures of this property were proposed \cite{P2,P3}. However, in order to avoid the complications encountered in extremization procedures of the previously defined distance-type degrees of entanglement \cite{Pl}, another pattern was followed in the case of non-Gaussianity degrees \cite{P2,P3}. Let us denote by ${\hat \rho}_G$ the Gaussian state having the same average displacement and covariance matrix as the given state 
$\hat \rho$. This state was reasonably used as a reference one in defining a degree of non-Gaussianity as the distance between the given state and the Gaussian state ${\hat \rho}_G$. In Refs.\cite{P2,P3,P33}, Genoni {\em et al.} employed the Hilbert-Schmidt metric to define a corresponding non-Gaussianity measure,
\begin{equation}
\delta_{\mathrm{HS}}[\hat \rho]:=\frac{{d}^2_{\mathrm{HS}}
(\hat \rho, {\hat \rho}_G)}{2{\rm Tr} (\hat \rho^2)}
=\frac{1}{2}\left[1+\frac{{\rm Tr } (\hat \rho_G^2)-2 {\rm Tr} (\hat \rho_G \hat \rho)}{{\rm Tr } (\hat \rho^2)}\right] ,
\label{hs}
\end{equation}
and the relative entropy to introduce the entropic measure 
of non-Gaussianity:
\begin{equation}
\delta_{\mathrm{RE}}[\hat \rho]:={\cal S}(\hat \rho|{\hat \rho}_G).
\label{re}
\end{equation}
On the one hand, Eq.\ (\ref{hs}) gives an easily computable expression. On the other hand, despite its not being a true distance, the relative entropy is acceptable and used 
as a measure of distinguishability between two quantum states. 
It is remarkable that the relative entropy\ (\ref{re}) reduces 
to a difference of von Neumann entropies:
\begin{equation} 
\delta_{\mathrm{RE}}[\hat \rho]={\cal S}(\hat \rho_G)
-{\cal S}(\hat \rho),
\label{re1}
\end{equation}
where ${\cal S}(\hat \rho):=-{\rm Tr}(\hat \rho\ln {\hat \rho})$ 
is the von Neumann entropy of the state $\hat \rho$.
General properties of the non-Gaussianity measures \eqref{hs} and  \eqref{re1} were discussed in detail in Refs.\cite{P2,P3,P33}.
Another approach to quantifying non-Gaussianity is based on the $Q$ function  and leads to a measure expressed by the difference between the Wehrl entropies of the Gaussian state ${\hat \rho}_G$ and the given non-Gaussian state $\hat \rho$ \cite{Simon}.
It is also worth noting that the non-Gaussianity measures \cite{P2,P3} were used to check on the relation between non-Gaussianity and the possibility of extending Hudson's theorem to  mixed states  \cite{Cerf2,Cerf3}.  In the two-mode case, the relation between 
non-Gaussianity  and entanglement both  quantified by the relative entropy was analyzed for photon-added and photon-subtracted two-mode squeezed vacuum states \cite{Cerf4}. From the experimental part, non-Gaussianity in terms of relative entropy \cite{P3} was measured for single-photon-added coherent states \cite{P4}. 
An experimental study on the same measure was recently reported 
for phase-averaged coherent states \cite{Bondani}.  

The aim of the present paper is threefold. First, we reinforce 
the Bures distance to the associate Gaussian state 
${\hat \rho}_G$ as a measure of non-Gaussianity of an arbitrary 
one-mode state $\hat \rho$. We have recently introduced it on general grounds in Ref.\cite{GMM} by using this well-known metric related to the fidelity between two quantum states \cite{Uhl}. 
Our motivation to use this measure was given by the valuable distinguishability properties of the fidelity described in Refs.\cite{Jo,BZ}. Application of this measure for $M$-photon-added thermal states showed us its good agreement with previously defined distance-type measures \cite{P2,P3,P33}. Note that fidelity-based metrics have proven to be fruitful in quantum optics and quantum information as measures of nonclassicality \cite{PTH02} and entanglement \cite{PTH,PT08a,PT08b}. Second, we use the three 
above-mentioned distance-type measures to investigate the loss 
of non-Gaussianity for $M$-photon-added thermal states of a field coupled to  a heat bath. To this end, we give a new derivation 
of the Fock-basis solution of the quantum optical master equation by employing very conveniently the Weyl expansion of the density operator.  Third, we write the damped quasiprobability densities and examine their evolution towards the equilibrium thermal state imposed by the reservoir. This evolution is marked by the loss 
of negativity of both Wigner and $P$ functions. Collaterally, 
we notice the nonmonotonic evolution of the von Neumann entropy during decoherence as depending on the mean photon number of the input state.

The plan of the paper is as follows. In Sec. II  we recapitulate the principal features of the Bures degree of non-Gaussianity arising from some general properties of the fidelity. We here concentrate on an easily workable case, namely, that of mixed 
states having diagonal density matrix in the Fock basis. 
A comparison between various degrees of non-Gaussianity is further intended. Therefore, we consider in Sec. III a set of states generated as solutions of the quantum optical master equation 
\cite{BP}. The density matrix of the damped field mode and the quasiprobability densities at the moment $t$ are derived for an arbitrary input state which is diagonal in the Fock basis. The evolution of the quasiprobability densities  allows us the formulation of some general properties.
Section IV is dedicated to a specific damped state which is important in applications: An $M$-photon-added thermal state evolving under the influence of a thermal reservoir. With this explicit input density matrix  at the moment $t=0$ and exploiting  some results of Sec. III, we here find the damped quasiprobability distributions and density matrix  in a closed analytic form. Further, the time evolutions of the Hilbert-Schmidt, entropic, 
and Bures measures of non-Gaussianity are analyzed and compared.  Special attention is paid to the entropy production 
${\cal S}(\hat\rho)$ as another tool to examine Gaussification 
of single-mode states due to the field-reservoir interaction. Section V summarizes our results focused on the consistency 
of the analyzed measures of non-Gaussianity and the principal features of nonclassicality decay considered in the present paper.

\section{Bures measure of non-Gaussianity}

Let us consider two arbitrary states, 
$\hat{\rho}$ and $\hat{\sigma}$, of a given quantum system. According to Uhlmann, when both states are mixed, a good measure
of the closeness between their properties is the maximal 
quantum-mechanical transition probability between their purifications in an enlarged Hilbert space \cite{Uhl,Jo}. 
This is an extended notion of transition probability between quantum states which is now called fidelity \cite{Jo} 
and has the intrinsic expression \cite{Uhl}
\begin{equation}
{\cal F}(\hat\rho, \hat\sigma )=\left\{ {\rm Tr}\left[ \left(
\sqrt{\hat\rho}\hat\sigma \sqrt{\hat\rho}\right)
^{1/2}\right] \right\}^2.
\label{F} 
\end{equation}
In Ref.\cite{GMM} we have defined a fidelity-based degree 
of non-Gaussianity 
\begin{equation} 
\delta_{F}[\hat \rho]:=\frac{1}{2}\;{d}^2_{B}(\hat \rho, 
{\hat \rho}_G)=1-\sqrt{{\cal F}(\hat \rho,\hat \rho_G )}. 
\label{bu}
\end{equation}
As seen in Eq.\ (\ref{bu}), the fidelity is tightly related to 
the Bures metric ${d}_{B}$  introduced in Ref.\cite{Bu} 
on mathematical grounds. Several general properties of fidelity 
are listed below together with the features they transfer 
to our definition\ (\ref{bu}). 

{\bf (P1)}  $0\leqq {\cal F}(\hat \rho, \hat \sigma)\leqq 1,\;$ 
and\; ${\cal F}(\hat \rho, \hat \sigma)=1\;$ if and only if 
$\hat \rho=\hat \sigma$. This property implies
\begin{equation}
\delta_{F}[\hat \rho]=0, \quad {\rm iff} \;\; \hat \rho\;\; 
{\rm is}\;\; {\rm Gaussian},
\label{p1}
\end{equation}
\begin{equation}
0<\delta_{F}[\hat \rho] \leqq 1,\quad {\rm otherwise}.
\label{p11}
\end{equation}
Properties\ (\ref{p1}) and\ (\ref{p11}) justify 
the interpretation of the quantity $\delta_{F}[\hat \rho]$ 
as a degree of non-Gaussianity. 

{\bf (P2)}   ${\cal F}(\hat \sigma, \hat \rho)={\cal F}(\hat \rho, \hat \sigma) \quad$ (symmetry).

{\bf (P3)}   ${\cal F}(\hat \rho, \hat \sigma)\geqq {\rm Tr} 
(\hat \rho \,\hat \sigma).\,$  When at least one of the states 
is pure, then the inequality is saturated:
${\cal F}(\hat \rho, \hat \sigma)
={\rm Tr} (\hat \rho \,\hat \sigma).$ 
Accordingly, for any pure state $|\Psi\rangle \langle \Psi|,$ 
Eq.\ (\ref{bu}) reads
\begin{equation}
\delta_{F}[|\Psi\rangle\langle \Psi|]= 1-\sqrt{\langle \Psi|
\hat{\rho}_G|\Psi\rangle}.
\label{pure}
\end{equation}
Unlike $\delta_{\mathrm{RE}}$, which has the same value for all pure states possessing the same covariance matrix, $\delta_{F}$ 
is state dependent.

{\bf (P4)} ${\cal F}(\hat U \hat \rho \hat U^{\dag}, \hat U \hat \sigma\hat U^{\dag})=
{\cal F}(\hat \rho, \hat \sigma) \;$ (invariance under unitary transformations). As shown in Refs.\cite{P2,P3,P33}, when $\hat U$ are the unitary operators  of the metaplectic representation 
on the Hilbert space of states, then the property 
$$ \hat \rho^{\prime}=\hat U \hat \rho \hat U^{\dag} \Longrightarrow (\hat \rho^{\prime})_G=
\hat U \hat \rho_G \hat U^{\dag}$$ 
holds, and therefore
\begin{equation}
\delta_{F}[\hat U \hat \rho \hat U^{\dag}]=\delta_{F}[\hat \rho].
\label{p2}
\end{equation}
It follows that, in the one-mode case, $\delta_{F}[\hat \rho]$ 
does not depend on one-mode squeezing and displacement operations. 

{\bf (P5)} ${\cal F}(\Phi(\hat \rho), \Phi(\hat \sigma))\geqq 
{\cal F}(\hat \rho, \hat \sigma)$ (monotonicity under any 
trace-preserving, convex-linear, and completely positive map 
$\Phi$). This means that fidelity does not decrease strictly 
under any trace-preserving quantum operation, including an arbitrary nonselective measurement. Therefore, by virtue of definition\ (\ref{bu}), the Bures degree of non-Gaussianity does
not increase under any such quantum operation. 

{\bf (P6)} ${\cal F}(\hat \rho_1 \otimes \hat \rho_2, 
\hat \sigma_1 \otimes \hat \sigma_2)=
{\cal F}(\hat \rho_1, \hat \sigma_1){\cal F}(\hat \rho_2, 
\hat \sigma_2)\;$ (multiplicativity). Let us consider a bipartite product state $\hat \rho:=\hat \rho_1\otimes \hat \rho_2$. If  $\hat \rho_2$ is a Gaussian state, we get 
$${\cal F}(\hat \rho_1 \otimes \hat \rho_2, (\hat \rho_1)_G 
\otimes \hat \rho_2)={\cal F}(\hat \rho_1, (\hat \rho_1)_G)$$ and therefore
\begin{equation}\delta_{F}[\hat \rho_1 \otimes \hat \rho_2]
=\delta_{F}[\hat \rho_1].
\label{p3}
\end{equation}

{\bf (P7)} For commuting density operators, 
$\; [\hat \rho,\hat \sigma]=\hat 0, \;$
Eq.\ (\ref{F}) simplifies to 
\begin{equation}
{\cal F}(\hat \rho, \hat \sigma)=\left[ {\rm Tr}\left( 
\sqrt{\hat \rho}\sqrt{\hat \sigma}\right) \right] ^2.
\label{p4}
\end{equation}

Properties \ (\ref{p1}),\ (\ref{p2}), and\ (\ref{p3}) 
of $\delta_{F}[\hat \rho] $ are shared by the non-Gaussianity measures\ (\ref{hs}) and\ (\ref{re1}) as well \cite{P33}.
In general, the Hilbert-Schmidt measure\ (\ref{hs}) is easier to compute than $\delta_{F}[\hat \rho] $ or $\delta_{\mathrm{RE}}
[\hat \rho]$ and one could wonder why these last two were in fact considered?  The answer is given by the monotonicity property 
under any trace-preserving quantum operation shared by the relative entropy and the fidelity, unlike the Hilbert-Schmidt distance 
\cite{BZ}. One of our goals in this work is to notice the role 
the monotonicity plays in the evolution of non-Gaussianity under damping (see Sec. IV).

In the following we concentrate on an interesting computable case. 
The Gaussian reference state of a Fock-diagonal non-Gaussian one,
\begin{equation}
\hat \rho=\sum_{l=0}^{\infty} p_l \, |l\rangle \langle l| \quad
{\rm with} \quad  \sum_{l=0}^{\infty}  p_l=1, 
\label{diag}
\end{equation}
is a thermal state with the same mean photon occupancy 
$\langle \hat N \rangle=\sum_l l\,p_l$:
\begin{equation}
\hat\rho_G=\sum_{l=0}^{\infty} s_l |l\rangle \langle l| \quad
{\rm with} \quad s_l:=\frac{\langle \hat N \rangle^l}
{(\langle \hat N \rangle+1)^{l+1}}.
\label{diag1}
\end{equation}
The corresponding Hilbert-Schmidt and entropic degrees of non-Gaussianity were written in Refs.\cite{P2,P3} as:
\begin{align}
\delta_{\mathrm{HS}}[\hat \rho]=\frac{1}{2}\left[1+\frac{1}
{\sum_{l}p_l^2}\left(\frac{1}{2\langle \hat N 
\rangle+1}-2\sum_{l}p_l s_l\right) \right], 
\notag\\
\label{hs2}
\end{align}
where ${\rm Tr} (\hat\rho_G^2)=(2\langle \hat N 
\rangle+1)^{-1}$ is the purity of the state $\hat\rho_G$,
and
\begin{align}
\delta_{\mathrm{RE}}[\hat \rho]=\sum_{l}p_l\ln{p_l}+
(\langle \hat N \rangle+1)\ln (\langle \hat N \rangle+1)-\langle \hat N \rangle\ln \langle \hat N \rangle.
\notag\\
\label{re2}
\end{align}
Here we have employed the von Neumann entropy 
${\cal S}(\hat{\rho}_G)$ of a Gaussian state. In this special case, we notice the commutation relation $[\hat\rho,\hat{\rho}_G]=\hat0$,
which allows one the use of Eq.\ (\ref{p4}) to get
\begin{equation}
\delta_{F}[\hat\rho]=1-\sum_{l=0}^{\infty} \sqrt{ p_l\, s_l}.
\label{diag2}
\end{equation}
An important example of states having the density operator 
of the form\ (\ref{diag}) is the class of phase-averaged 
(or phase-randomized) coherent states. A complete theoretical 
and experimental characterization of such states including 
measurements of the degrees of non-Gaussianity\ (\ref{hs2}),
\ (\ref{re2}), and\ (\ref{diag2}) was recently reported 
in Ref.\cite{OL2013}. Note finally that various
excitations on a thermal state of the type 
$\hat\rho \sim (\hat a^{\dag})^k \;\hat a^l\;\hat 
\rho_{\mathrm{th}}\; (\hat a^{\dag})^l \;\hat a^k $ 
have the density operator of the type\ (\ref{diag}). Here $\hat a$ and $\hat a^{\dag}$ are the amplitude operators of the field mode.

\section{Master equation of one-mode field damping}

Loss of nonclassical properties of the single-mode field states (Gaussian and non-Gaussian as well) during the interaction with 
a dissipative environment has been intensively studied in the last decades \cite{Loss,Mil,Buz,PT93b,Muss,PT96}. Quite recently, more general  properties of damped non-Gaussian states such as the evolution of mixing measured by linear-entropy production 
\cite{Isar,PT00a,PT00b}, or the evolution of various measures 
of nonclassicality \cite{P2011} were investigated by employing 
the quantum optical master equation in the interaction 
picture \cite{BP}: 
\begin{align}
\frac{\partial\hat \rho}{\partial t} & =
\frac{\gamma}{2}(2\hat a\hat \rho\hat  a^{\dag}-\hat a^{\dag}
\hat a\hat \rho-\hat \rho \hat a^{\dag}\hat a)
\notag \\
& +\gamma \bar{n}_R (\hat a^{\dag}\hat \rho \hat a+
\hat a\hat \rho \hat a^{\dag}-\hat a^{\dag} \hat a \hat \rho-\hat \rho \hat a \hat a^{\dag}).
\label{me}
\end{align}
In Eq.\ (\ref{me}), $\hat \rho$ is the reduced density operator 
of the field, $\gamma$ is the coupling constant between field 
and bath, and ${\bar n}_R$ stands for the mean occupancy of the reservoir. This  popular master equation is of the Lindblad type and thus preserves both the positivity and normalization of the density operator. It has the clear physical significance of describing decoherence of a field mode coupled to a heat bath. 
For a comprehensive list of references regarding solutions arising from Eq.\ (\ref{me}) for the expectation values of field operators, characteristic functions, and quasiprobability densities, we refer the reader to some recent work of Dodonov \cite{Dod1,Dod2}. 

\subsection{Damped density matrix}

We here intend to write the evolving damped-mode state generated 
by this master equation from an input non-Gaussian one. Then 
we will compare the time evolution of the above-discussed degrees of non-Gaussianity. To go on this programme, we need a workable solution of Eq.\ (\ref{me}) in the Fock basis. In what follows 
our main tool is the Weyl expansion of the density operator: 
\begin{equation} 
\hat \rho=\frac{1}{\pi}\int d^2{\lambda}\, \chi(\lambda)\, 
\hat D(-\lambda).
\label{2.1bis}
\end{equation} 
In Eq.\ (\ref{2.1bis}), $\chi(\lambda)$ is  the characteristic function (CF) of the state $\hat \rho$, defined as the expectation value of the  displacement operator 
$\hat D(\lambda)=\exp(\lambda \hat a^{\dag}
-\lambda^{*} \hat a)$,
\begin{equation} 
\chi(\lambda):={\rm Tr} [\hat \rho \hat D(\lambda)].
\label{cf}
\end{equation} 
The CF $\chi(\lambda, t)$ of the damped field state is found 
to be determined by its initial form $\chi(\lambda, 0)$ 
\cite{Rock,PT00a,PT00b}:
\begin{align}
\chi(\lambda,t) & =\chi \left(\lambda \, e^{-\frac{\gamma}{2}t}, 
0 \right )
\notag \\
& \times \exp{\left[ -\left( \bar{n}_R+\frac{1}{2}\right)
(1-e^{-\gamma t})|\lambda|^2\right] }.
\label{4}
\end{align}
The mean photon number in the damped mode is 
\begin{equation}
\langle \hat{N}\rangle {\vert}_t=\langle \hat{N}\rangle {\vert}_0
\, e^{-\gamma t}+{\bar n}_{T}(t),
\label{num}
\end{equation} where
${\bar n}_{T}(t):={\bar n}_R(1-e^{-\gamma t})$
denotes the thermal mean occupancy in the field mode at time $t$ 
and $\langle \hat{N}\rangle {\vert}_0$ is the initial mean photon number.

We expect that the damping master equation \eqref{me}, which is 
phase insensitive, preserves the diagonal form of an evolving 
state whose initial density matrix is Fock-diagonal. Indeed, 
the CF of an input Fock-diagonal state $\hat \rho (0)$ is 
\begin{equation}
\chi(\lambda,0)=\exp{\left(-\frac{1}{2}|\lambda|^2 \right)}
\sum_{l=0}^{\infty} \rho_{ll}(0)\;L_l(|\lambda|^2),
\label{cf0}
\end{equation} 
where $\rho_{ll}(0)$ is the photon-number distribution 
and $L_{l}(x)$ is a Laguerre polynomial of degree $l$.
Equation\ (\ref{4}) gives further 
\begin{align}
\chi(\lambda,t) & =\exp{\left\{ -\left[ {\bar n}_{T}(t)
+\frac{1}{2}\right] |\lambda|^2\right\}}
\notag \\
& \times \sum_{l=0}^{\infty}\, \rho_{ll}(0)\;
L_l{\left(\, |\lambda|^{2}e^{-\gamma t}\,\right)}.
\label{cft}
\end{align}
The density matrix $\rho_{jk}(t)$ of the damped field state is
obtained from the Weyl expansion\ (\ref{2.1bis}) after here inserting the CF\ (\ref{cft}).  Use is also made of the matrix elements of the displacement operator in the Fock basis. After a simple calculation using the polar coordinates and an obvious change of variable in the integral\ (\ref{2.1bis}) we are left 
with the following series expansion:
\begin{align}
\rho_{jk}(t) & =\delta_{jk}\sum_{l=0}^{\infty}\rho_{ll}(0)
\notag \\
& \times \int_{0}^{\infty} dx\, e^{-[{\bar n}_{T}(t)+1]x}\,
L_l(x{e}^{-\gamma t})\, L_j(x).
\label{dm}
\end{align}
The integral in the above equation can be routinely performed 
\cite{RG} yielding the formula
\begin{align}
& \rho_{jk}(t)=\delta_{jk}\frac{[{\bar n}_{T}(t)]^j}
{[{\bar n}_{T}(t)+1]^{j+1}}\, \sum_{l=0}^{\infty}\rho_{ll}(0)
\notag \\
& \times\left[\frac{({\bar n}_R+1)
(1-e^{-\gamma t})}{{\bar n}_{T}(t)+1}\right]^l 
\notag \\ 
& \times {_2F_1}\left[ -j, -l;1;\frac{e^{-\gamma t}}
{({\bar n}_R+1)(1-e^{-\gamma t}){\bar n}_{T}(t)}\right] ,
\label{dm1}
\end{align} 
where ${_{2}F_{1}}$ is a Gauss hypergeometric function, 
Eq.\ (\ref{a1}).  Equation\ (\ref{dm1}) is the general solution for the damped density matrix as function of the input one 
in the Fock-diagonal case. Let us take the limit $t\rightarrow \infty$ in Eq.\ (\ref{dm1}). The result is a thermal state with the 
Bose-Einstein mean occupancy $\bar{n}_R$. We thus deal with 
a {\em Gaussification process}, namely, evolution under this master equation eventually destroys the non-Gaussianity and also the 
nonclassicality properties of any input state. A special case of 
Eq.\ (\ref{dm1}) arises for the thermal contact with a zero-temperature bath ($\bar n_R=0$), i.e., for the field coupling to the vacuum. The evolving state \ (\ref{dm1}) then becomes:
\begin{align}
\rho_{j k}(t)=\delta_{jk}\sum_{l=j}^{\infty}\,
\binom{l}{j}\, \rho_{ll}(0)\; e^{-j\gamma t}
\left( 1-e^{-\gamma t}\right)^{l-j}.
\label{dm2}
\end{align} 
Equation\ (\ref{dm2}) describes both dissipation by contact 
with a zero-temperature reservoir and photon counting for which 
the exponential $e^{-\gamma t}$ should be simply replaced
by the quantum efficiency $\eta$ of the detector \cite{PT93b}. 

\subsection{ Damped quasiprobability densities}

We recall the most useful $s$-ordered CFs \cite{CG}: 
\begin{align}
\chi(\lambda;s): & =\exp{\left(\frac{s}{2}\,|\lambda|^2 \right)}\;
{\rm Tr} [\hat{\rho}\, \hat{D}(\lambda)],
\notag\\
& (s=-1,0,1).
\label{cf-gen}
\end{align}
Their corresponding normalized Fourier transforms 
\begin{equation}
W(\beta;s)=\frac{1}{\pi^2}\int d^2 \lambda \,
\exp{(\beta \lambda^*-\beta^* \lambda)} \;\chi(\lambda;s)
\label{qd-gen}
\end{equation}
can be interpreted as quasiprobability densities and are important tools for understanding the nonclassical features of quantum states. Specifically, they are: for $ s=1$, $P(\beta):=W(\beta;1)$ (Glauber's $P$ function), $s=0$, $W(\beta):=W(\beta;0)$ 
(Wigner function), and $ s=-1$, $Q(\beta):=W(\beta;-1)$ (Husimi function).

Our aim here is to write a general expression for the 
quasiprobability densities  describing a damped
state. Therefore we insert the CF \eqref{cft} 
into Eq.\ (\ref{qd-gen}) and get 
\begin{align}
& W(\beta,t;s)=\frac{1}{\pi^2}\sum_{l=0}^{\infty} \rho_{ll}(0)
\int d^2\lambda \, \exp{(\beta \lambda^*-\beta^* \lambda)}
\notag \\ 
& \times\exp{\left\{-\left[{\bar n}_{T}(t)+\frac{1}{2}(1-s)
\right] |\lambda|^2\right\}}\, L_l(|\lambda|^2\, e^{-\gamma t}).
\label{qd}
\end{align}
We first perform the integration over the polar angle, 
then we are left with a known integral over $|\lambda|$ whose evaluation \cite{RG} leads us to the following double summation:
\begin{align}
W(\beta,t;s) & =\frac{1}{\pi}\sum_{l=0}^{\infty}\, \rho_{ll}(0)\; 
\frac{[A(t;s)]^l}{\left[ {\bar n}_{T}(t)
+\frac{1}{2}(1-s)\right]^{l+1}}
\notag \\ 
& \times \sum_{k=0}^{\infty}\frac{(-1)^k |\beta|^{2 k}}{k! \left[{\bar n}_{T}(t)+\frac{1}{2}(1-s)\right]^k}
\notag \\ 
&\times {_2F_1}\left[ -k, -l;1;\,-\frac{e^{-\gamma t}}
{A(t;s)}\right].
\label{qd1}
\end{align}
In Eq.\ (\ref{qd1}) we have introduced a bath-dependent parameter
whose relevance for the evolution of the quasiprobability densities will arise further:
\begin{align}
A(t;s) &:=\bar n_{T}(t)+\frac{1}{2}(1-s)-e^{-\gamma t}
\notag \\
& =\bar n_R+\frac{1}{2}(1-s)-(\bar{n}_R+1)\, e^{-\gamma t}.
\label{a}
\end{align}
Now, the sum over $k$ is of the type\ (\ref{a55}) and leads
to a rather simple result:
\begin{align}
W(\beta,t;s) & =\frac{1}{\pi}\exp{\left[ -\frac{|\beta|^{2}}
{{\bar n}_{T}(t)+\frac{1}{2}(1-s)}\right] }
\notag \\ 
& \times \sum_{l=0}^{\infty} \rho_{ll}(0)\; \frac{[A(t;s)]^l}
{\left[ {\bar n}_{T}(t)+\frac{1}{2}(1-s)\right]^{l+1}}
\notag \\ 
& \times L_l\left\{ -\frac{|\beta|^{2}\, e^{-\gamma t}}
{[{\bar n}_{T}(t)+\frac{1}{2}(1-s)]A(t;s)}\right\} .
\label{qd2}
\end{align}
Any Laguerre polynomial $L_l(x)$ displays, via Eqs.\ (\ref{lag}) 
and\ (\ref{a11}), its positivity for negative values 
of the argument $x$. Equation\ (\ref{qd2}) presents two main advantages. First, it makes possible a simultaneous evaluation of all quasiprobability densities $W(\beta,t;s)$. Second, its structure implies a general statement regarding the positivity 
of the quasiprobability densities independent of the initial state 
$\hat \rho(0)$. Indeed, at any $\beta$ all the quasiprobability densities $W(\beta,t;s)$ [Eq.\ (\ref{qd2})] are positive 
for positive $A(t;s)$. This allows us to write the ultimate times at which any quasiprobability distribution displaying initially some negativies becomes positive due to the field interaction 
with the thermal bath. 

(i) $s=-1 \Rightarrow A(t;-1)=(\bar n_R+1)(1-e^{-\gamma t})>0$.
As expected from its definition as the average value of the density operator in a coherent state, the $Q$ function is always positive.

(ii) $s=0 \Rightarrow  A(t;0)=\bar n_R+\frac{1}{2}-(\bar n_R+1)\,
e^{-\gamma t}$ is positive for
\begin{equation}
t\geqq t_w:=\frac{1}{\gamma}\ln \left( 1+\frac{1}{2\bar n_R+1}\right).
\label{tw}
\end{equation}
This is the time at which the Wigner function completely loses 
any negativity.

(iii) $s=1 \Rightarrow A(t;1)=\bar n_R-(\bar n_R+1)\,
e^{-\gamma t}$
is positive for 
\begin{equation}
t\geqq t_c:=\frac{1}{\gamma}\ln \left(1+\frac{1}{\bar n_R}\right).\label{tc}
\end{equation}
Beyond the time $t_c$, the Glauber-Sudarshan $P$ representation exists as a genuine probability density. The threshold 
times\ (\ref{tw}) and\ (\ref{tc}) are constants of the bath 
valid for any input Fock-diagonal state.  Note also that $t_c>t_w$,
that is, the Wigner function is more fragile than the $P$ function under damping. Let us now remark that the characteristic time
\ (\ref{tc}) was previously obtained for particular input states such as even coherent states in Refs.\cite{PT00b,P2011} 
and Fock states in \cite{PT00a}. The limit time\ (\ref{tw}) 
was found in Refs.\cite{PT00a,PT00b} in describing the mixing process during the damping of an input pure state. As an onset of positivity of the Wigner function for damped photon-added thermal states, the time $t_w$ was written in Refs.\cite{li,shu}. In the case of a zero-temperature reservoir, $\bar n_R=0$, we get an unique state-independent time $t_w=\frac{1}{\gamma}\ln 2$ for disappearance of any negativity of the Wigner function, while, according to Eq.\ (\ref{tc}), the $P$ representation could have negative domains at any time.

What happens at earlier times $t<t_w$? We cannot make general predictions using our Eq.\ (\ref{qd2}). The behavior of the distribution functions $P(\beta, t)$ and $W(\beta, t)$ will depend on the initial state $\hat \rho(0)$. 

\section{A case study: damped photon-added thermal states}
 
According to a pertinent remark of Lee \cite{Lee}, adding photons to a classical state implies removal of the vacuum state from its Fock expansion. A nonclassical output is obtained in every case. Addition of photons to a classical Gaussian state, such as a coherent or a thermal one, generates a non-classical output which is no longer Gaussian. We can say that photon-added states are both non-classical and non-Gaussian. 
An $M$-photon-added thermal state \cite{AT,JL} is an interesting example of a Fock-diagonal state whose non-classicality was recently investigated experimentally \cite{Bel,Kies08, Kies11}. 
The density operator of the state obtained by adding $M$ photons 
to a thermal one $\hat \rho_{\mathrm{th}}$ is
\begin{equation}
\hat \rho^{(M)}=\frac{1}{M!\, (\bar n+1)^M}\, 
(\hat a^\dagger )^M\, \hat \rho_{\mathrm{th}}\, \hat a^M.
\label{PATS}\end{equation}
The spectral decomposition of a given thermal state 
$\hat{\rho}_{\mathrm{th}}$ is of the type\ (\ref{diag1}), 
with a mean photon number denoted  $\bar n$. The nonclassicality 
of the state\ (\ref{PATS}) was first discussed by Agarwal and Tara \cite{AT} in terms of its nonpositive $P$ representation 
and Mandel's $Q$-parameter. The non-Gaussianity 
of the state\ (\ref{PATS}) was recently evaluated 
in Ref.\cite{Simon} by employing the Wehrl-entropy measure and found to coincide with the non-Gaussianity of the number state 
$|M\rangle\langle M|$, being thus independent of the thermal mean occupancy $\bar n$. We employ here Eq.\ (\ref{PATS}) to write down the density matrix:
\begin{equation}
(\hat \rho^{(M)})_{lm}=\delta_{lm}\, \binom{l}{M}
\frac{{\bar n}^{l-M}}{(\bar n+1)^{l+1}}=:p_l\delta_{lm}.
\label{pl}
\end{equation}
The photon-number distribution of the associate thermal state 
\begin{equation}
(\hat \rho_G^{(M)})_{ll}=\frac{\left[\bar n (M+1)+M\right]^l}
{[(M+1)(\bar n+1)]^{l+1}}=:s_l
\label{sl}
\end{equation}
is obtained by using the shared mean photon occupancy  
$\langle \hat N\rangle=\bar{n} (M+1)+M$. The non-Gaussianity 
degrees\ (\ref{hs2}),\ (\ref{re2}), and\ (\ref{diag2}) 
for the state\ (\ref{PATS})  were carefully analyzed in our work \cite{GMM}. We have there adopted as a reasonable criterion for their appropriateness a monotonic behavior with respect to the average photon number  $\langle \hat N\rangle$ of the state. 
In the case of Hilbert-Schmidt degree\ (\ref{hs2}) we were able to give an analytic result, while for the fidelity-based degree and the relative-entropy measure we performed numerical evaluations. 
We found that they depend on the thermal mean occupancy $\bar n$ unlike the  Wehrl-entropy measure \cite{Simon}. Our evaluations have also shown a consistent relation between these three 
non-Gaussianity measures. Their behavior under damping will be analyzed in the following by extending the results reported 
in Ref.\cite{GMM}. 

\subsection{Decay of negativities}

The damped photon-added thermal states offer a good test bed 
to examine the evolution of the initial quasiprobability distributions with negativities towards positive representations following our general findings in Sec. III B. We evaluate first 
the time development of quasiprobability densities 
$W(\beta, t; s)$ by inserting the density matrix\ (\ref{pl}) 
into Eq.\ (\ref{qd2}). By applying Eq.\ (\ref{la5}) we finally get after minor rearrangements:
\begin{align}
W(\beta, t;s) & =\frac{1}{\pi}\exp{\left[-\frac{|\beta|^{2}}
{B(t;s)}\right]} \frac{[A(t;s)]^M}{[B(t;s)]^{M+1}}
\notag \\
& \times L_M\left[-\frac{(\bar{n}+1)\, e^{-\gamma t}|\beta|^{2}}
{B(t;s)A(t;s)}\right] , 
\label{qdM}
\end{align}
where we have denoted 
\begin{equation}
B(t;s):=\bar n_{T}(t)+\frac{1}{2}(1-s)+\bar{n}\, e^{-\gamma t}, 
\label{b}
\end{equation}
and $A(t;s)$ is the universal bath parameter given 
by Eq.\ (\ref{a}). Let us now write the quasiprobability densities of the input state [Eq.\ (\ref{PATS})]. By specializing 
$\bar n_R=0,\,t=0$ in Eq.\ (\ref{qdM}), we easily get: 
\begin{equation}
Q(\beta,0)=\frac{1}{\pi}\frac{|\beta|^{2M}}{M!\, (\bar n+1)^{M+1}}\exp{\left(-\frac{|\beta|^{2}}{\bar n+1}\right)},
\label{q0}
\end{equation}
\begin{align}
P(\beta,0) & =\frac{1}{\pi}\frac{(-1)^M}{{\bar n}^{M+1}}
\, \exp{\left(-\frac{|\beta|^{2}}{\bar n}\right)}
\notag \\
& \times L_M\left(\frac{\bar{n}+1}{\bar n}|\beta|^{2}\right),
\label{p0}
\end{align}
\begin{align}
W(\beta,0) & =\frac{2}{\pi}\frac{(-1)^M}{(2\bar n+1)^{M+1}}
\, \exp{\left(-\frac{2|\beta|^{2}}{2\bar n+1}\right)}
\notag \\
& \times L_M\left[\frac{4(\bar n+1)|\beta|^{2}}{2\bar n+1}\right].
\label{w0}
\end{align}
We can see that the distributions $P(\beta,0)$ and $W(\beta,0)$ have a similar functional form which guarantees the existence of some negativity regions. Indeed, the Laguerre polynomial $L_M(x)$ 
has precisely $M$ distinct positive roots. According to 
Eq.\ (\ref{lag}), $L_M(0)=1$, so that $L_M(x)>0$ for $x \leqq 0$. On the contrary, a polynomial $L_M(x)$ is negative on a finite number of subintervals of $\mathbb{R}^+$: this number is either 
$\frac{M+1}{2}$, if $M$ is odd, or $\frac{M}{2}$, if $M$ is even. 
Therefore, a product of the type $(-1)^M L_M(x),\; x \geqq 0$, occurring in both expressions \eqref{p0} and \eqref{w0} presents negative values. Hence the $M$-photon-added thermal states 
are nonclassical for any values of the parameters $\bar n$ 
and $M$.

As regards the distributions of the damped field, we remark that the state-dependent quantity $B(t;s)$ is positive at any time. Therefore, the discussion on the characteristic times at which the distributions $P(\beta,t)$ and $W(\beta,t)$ lose their initial negativities as presented in Sec. III is equally valid here. 
There is no alteration of the limit times $t_c$ [Eq.\ (\ref{tc})] and $t_w$ [Eq.\ (\ref{tw})] produced by the specific input state [Eq.\ (\ref{PATS})] we are dealing with. The photon-added thermal states with their limit case $\bar n =0$, that is, a number state 
$|M\rangle\langle M|$, represent a perfect example of nonclassicality lost during Gaussification by damping. 
They develop in time through three distinct stages (regimes) 
of decoherence which can be related to their efficiency in continuous-variable quantum computation \cite{ME,VWFE,VFGE} 
as follows.

(1)  For $0\leqq t < t_w$, the damped states are nonclassical 
from both points of view of quantum optics (negative $P$ function) and quantum information processing (negative $W$ function) 
and could be used for quantum computational speed-up.

(2) For $t_w \leqq t < t_c$, the damped states have positive 
Wigner function, while their $P$ representation still displays negativities (region of bound universal states \cite{VFGE}). According to the recent results in Refs.\cite{ME,VWFE,VFGE}, 
such states admit an efficient classical simulation due to 
the positivity of their Wigner distribution, in spite of 
not having a well-behaved $P$ representation. 

(3) For $t\geqq t_c$, the damped states are classical and can be efficiently simulated on a classical computer. 

The discussion simplifies when dealing with a zero-temperature bath. We have $t_w=\frac{1}{\gamma}\ln 2$ and $t_c \rightarrow \infty$. This case  was recently tackled in Ref.\cite{VWFE} 
for $M=1$. The fidelity between the damped states and the vacuum was there considered as a measure of the closeness to the convex hull of Gaussian states. This was found to be small for $t>t_w$, namely, for the set of bound universal states created 
by dissipation. We shall further reconsider this problem 
for an arbitrary thermal reservoir by using the more rigourous measures of non-Gaussianity presented in Secs. I and II.

\subsection{Measures of non-Gaussianity}

 Insertion of the input photon-number distribution \eqref{pl} 
into Eq.\ (\ref{dm1}) gives the Fock-diagonal density matrix 
of a damped photon-added thermal state. Technically we have 
to evaluate the sum of a power series involving Gauss hypergeometric functions which is discussed in the Appendix.  
We apply Eq.\ (\ref{a5}) and get
\begin{align}
& {\rho^{(M)}}_{jk}(t)={\delta}_{jk}\left[ \frac{(\bar n_R+1)
(1-e^{-\gamma t})}{\bar{n}\, e^{-\gamma t}
+\bar n_{T}(t)+1}\right] ^{M}
\notag \\
& \times \frac{[\bar{n}\, e^{-\gamma t}+ \bar n_{T}(t)]^j}
{[\bar{n}\, e^{-\gamma t}+\bar n_{T}(t)+1]^{j+1}}
\notag \\
& \times {_2F_1}\bigg\{ -M,-j\, ;1;\, \frac{(\bar{n}+1)\,
e^{-\gamma t} }{(\bar n_R+1)(1-e^{-\gamma t})
[\bar{n}\, e^{-\gamma t}+{\bar n}_{T}(t)]}\bigg\}.
\notag \\
\label{dpa}
\end{align}
The reference Gaussian state associated with this density matrix 
is a thermal one whose mean occupancy equals the average photon number of the damped field state. According to Eq.\ (\ref{num}), this is
\begin{equation}
\langle \hat N \rangle |_t=\left[\bar n (M+1)+M\right] 
e^{-\gamma t}+{\bar n}_{T}(t).
\label{num1}
\end{equation}
We have used the time-dependent density matrix \eqref{dpa} 
and the mean occupancy \eqref{num1} to evaluate numerically 
two of the distance-type measures of non-Gaussianity we are interested in: $\delta_{F}[\hat \rho^{(M)}]$ and 
$\delta_{\mathrm{RE}}[\hat \rho^{(M)}]$ via Eqs.\ (\ref{diag2}) 
and \eqref{re2}, respectively. Our results are displayed in Figs. 1 
and 2, where the time evolutions of these non-Gaussianity measures are presented for weak damping ($\bar n_R=0.1$) and for a large 
thermal mean occupancy of the reservoir ($\bar n_R=5$), respectively. On both figures we have marked the three regions delimitated by the bath parameters $\gamma t_w$ and $\gamma t_c$.
In Fig. 1 the two regions of non-classicality are large enough 
($\gamma t_w=0.606$ and $\gamma t_c=2.398$). They contain states with higher non-Gaussianity, while the third (classical) region displays states slowly evolving to the Gaussian equilibrium state of the reservoir. Figure 2 shows that in a noisy bath 
the nonclassicality rapidly dissappears ($\gamma t_w=0.087$ 
and $\gamma t_c=0.182 $), while non-Gaussianity is still
large at the classicality threshold $\gamma t_c$. The functions 
$\delta_{F}[\hat \rho^{(M)}(t)]$ and 
$\delta_{\mathrm{RE}}[\hat \rho^{(M)}(t)]$ are monotonically decreasing, in accordance with the corresponding property 
discussed in Sec. II. Both Figs. 1 and 2 display a very good agreement between the two measures. They show consistency 
by inducing the same ordering of non-Gaussianity when considering specific sets of damped states continuously generated 
by the field-reservoir interaction.

\begin{figure*}[t]
\center
\includegraphics[width=7cm]{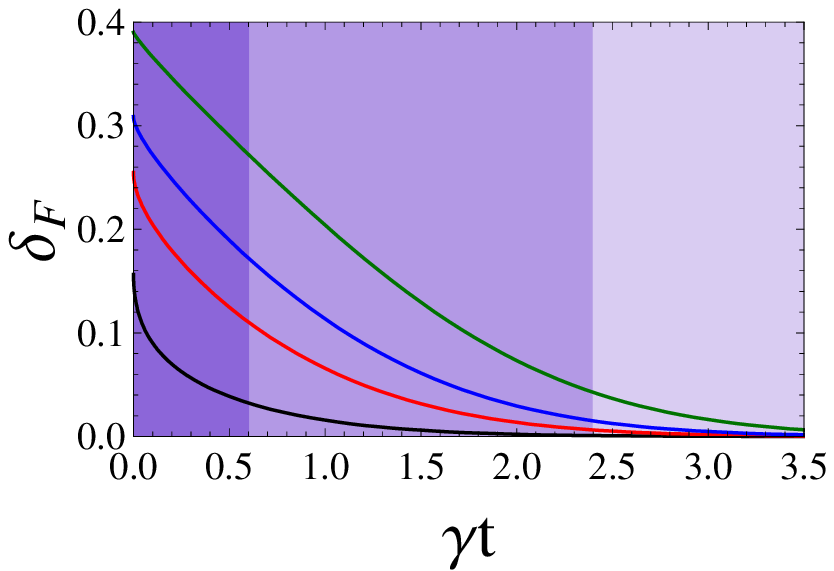}
\includegraphics[width=7cm]{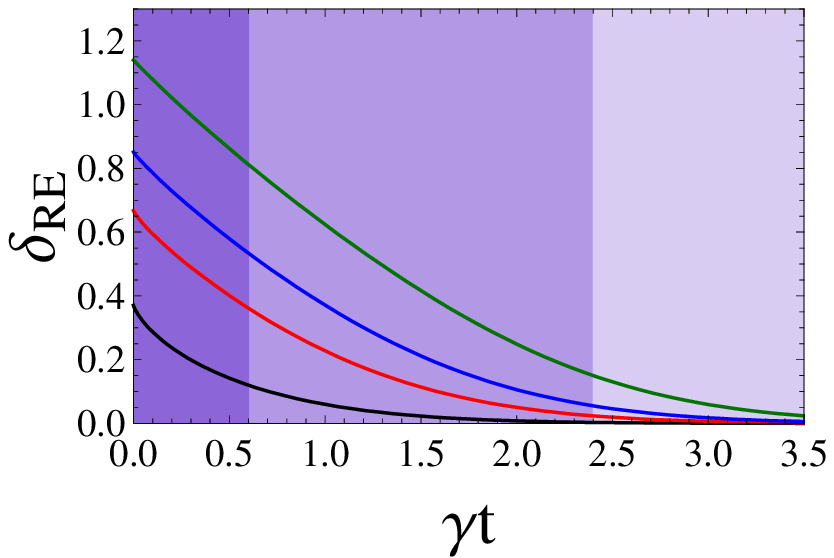}
\caption{Time evolution of the fidelity-based non-Gaussianity 
[Eq.\ (\ref{diag2})] (left plot) and the entropic measure 
[Eq.\ (\ref{re2})] (right plot), for input $M$-photon-added 
thermal states with $M=1,3,5,10$ (from bottom to top) 
and $\bar n=1$ in contact with a heat bath having $\bar n_R=0.1$. 
The nonclassicality threshold parameters are $\gamma t_w=0.606$ 
and $\gamma t_c=2.398$ and delimitate the three regions of decay.}
\label{fig-1}
\end{figure*}

\begin{figure*}[t]
\center
\includegraphics[width=7cm]{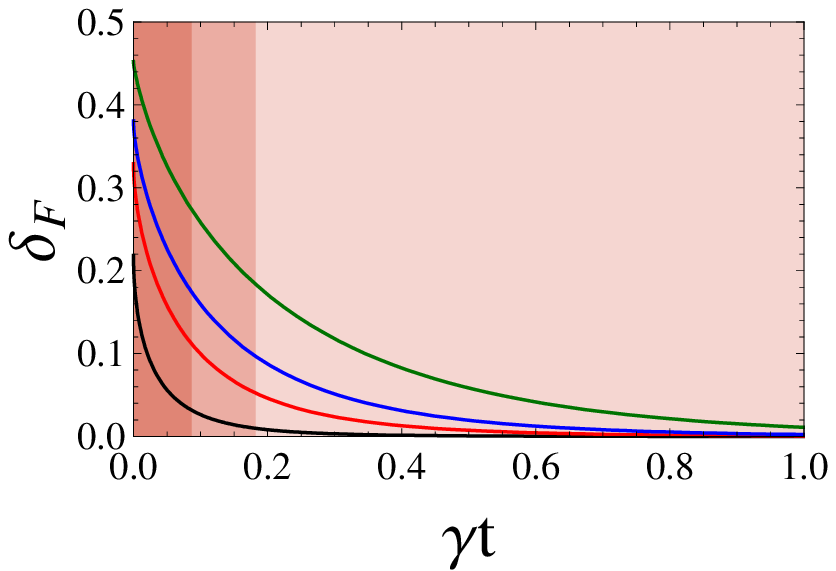}
\includegraphics[width=7cm]{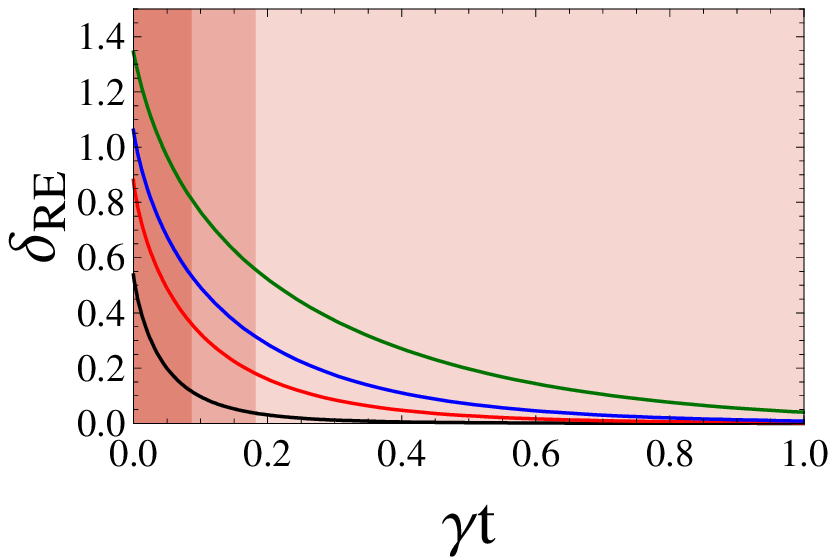}
\caption{As in Fig.1, but for input $M$-photon-added thermal states with $M=1,3,5,10$ (from bottom to top) and $\bar n=0.5$ in contact with a heat bath having $\bar n_R=5$. The characteristic parameters of this noisy bath are $\gamma t_w=0.087$ and $\gamma t_c=0.182 $.}
\label{fig-2}
\end{figure*}

The last degree of non-Gaussianity we are interested in 
is evaluated using Eq.\ (\ref{hs2}). Fortunately, 
as in the input-state case \cite{GMM}, we are able to give nice analytic expressions for both the scalar product 
${\rm Tr} [\hat \rho^{(M)}(t)\hat \rho^{(M)}_G(t)]$ 
and the  purity ${\rm Tr}\left\{ [\hat \rho^{(M)}(t)]^2\right\}$ 
of the damped state. We first write
\begin{align}
{\rm Tr} \left[\hat \rho^{(M)}(t) \hat \rho_G^{(M)}(t)\right]
& =\frac{1}{\langle\hat N \rangle |_t+1}\sum_{j=0}^{\infty}\,
\left( \frac{\langle \hat N \rangle |_t}
{\langle \hat N \rangle |_t +1}\right) ^j  
\notag \\
& \times \rho^{(M)}_{jj}(t),
\label{rg}
\end{align}
where the expectation value $\langle\hat N \rangle |_t$ 
is given in Eq.\ (\ref{num1}). By inserting Eq.\ (\ref{dpa}) 
for the damped density matrix into Eq.\ (\ref{rg}), 
we eventually get a sum of hypergeometric polynomials of the type 
\eqref{a3}. After a little algebra we obtain the formula
\begin{align}
& {\rm Tr} \left[ \hat \rho^{(M)}(t) \hat \rho_G^{(M)}(t)\right]
\notag \\
& =\frac{[\langle \hat N \rangle |_t+\bar n_{T}(t)+1
-e^{-\gamma t}]^M}{[\langle \hat N \rangle |_t
+\bar n_{T}(t)+\bar{n}\, e^{-\gamma t}+1]^{M+1}}. 
\label{rg1}
\end{align}
Now, the  purity of the damped state, 
\begin{align}
{\rm Tr} \left\{ \left[ \hat \rho^{(M)}(t)\right]^2\right\}
=\sum_{j=0}^{\infty}\left[\rho^{(M)}_{jj}(t)\right]^2,
\label{purity}
\end{align}
is proportional to the sum of a power series whose coefficients 
are squared Gauss hypergeometric polynomials. We apply 
the summation formula\ (\ref{a77}) and find after some minor rearrangements:
\begin{align}
& {\rm Tr} \left\{\left[\hat \rho^{(M)}(t)\right]^2\right\}
=\frac{[(\bar{n}+1)\, e^{-\gamma t}]^{2 M}}
{\left[ 2\bar n_{T}(t)+2\bar{n}\, e^{-\gamma t}+1\right] ^{2M+1}}
\notag \\
& \times{_2F_1}\bigg\{ -M,-M;\, 1;\, \frac{\left[ 2\bar n_{T}(t)
+ \bar{n}\, e^{-\gamma t}+1-e^{-\gamma t}\right] ^2}
{[(\bar{n}+1)\, e^{-\gamma t}]^2}\bigg\} .
\notag \\
\label{purity1}
\end{align}
Note that we can express the right-hand side 
of Eq.\ (\ref{purity1}) in terms of a Legendre polynomial 
via Eq.\ (\ref{a2}).  The equilibrium values of both the scalar product\ (\ref{rg1}) and the purity\ (\ref{purity1}) are equal to the purity of the thermal state eventually imposed 
by the reservoir, namely, $(2\bar n_R+1)^{-1}$. The interesting case of a number state $|M\rangle \langle M|$ is given by setting 
$\bar n=0$ in the above formulas. To compute the Hilbert-Schmidt degree of non-Gaussianity, we have to substitute 
into Eq.\ (\ref{hs2}) the overlap\ (\ref{rg1}), 
the purity\ (\ref{purity1}) of the damped state, and the purity 
of the reference Gaussian state,
\begin{equation}
{\rm Tr} \left\{ \left[ \hat \rho^{(M)}_G(t)\right] ^2\right\}
=\frac{1}{2 \langle \hat N \rangle |_t+1}.
\label{pg}
\end{equation}
In Fig. 3 we point out the time evolution of the Hilbert-Schmidt 
degree of non-Gaussianity for the same sets of states 
as in Figs. 1 and 2.
\begin{figure*}[t]
\center
\includegraphics[width=7cm]{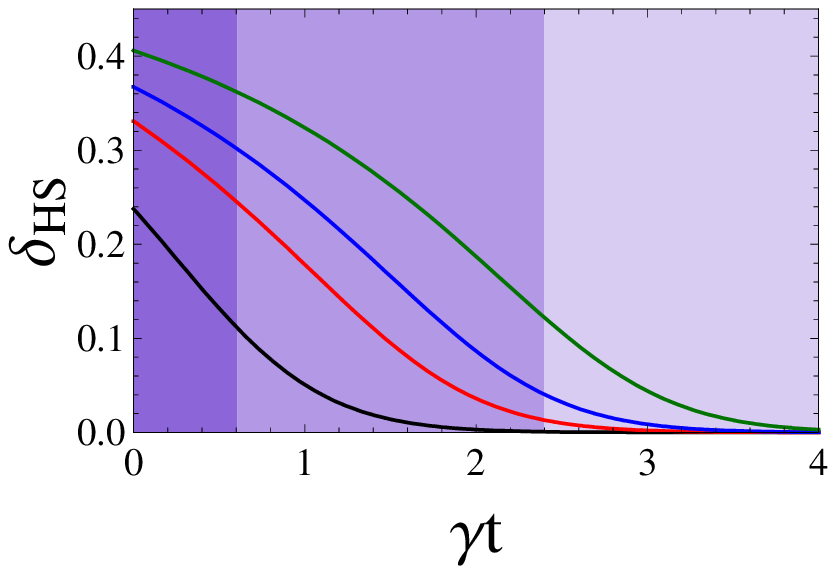}
\includegraphics[width=7cm]{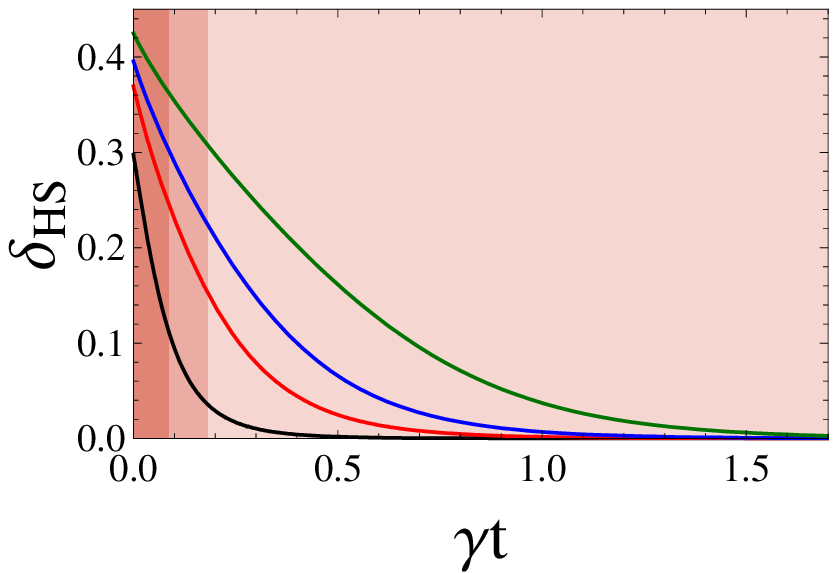}
\caption{Time evolution of the Hilbert-Schmidt degree of non-Gaussianity for contact with a heat bath when $\bar n_R=0.1$ 
(left plot) and $\bar n_R=5$ (right plot). The input states are 
$M$-photon-added thermal states with $\bar n=1$ (left) and  
$\bar n=0.5$ (right), and $M=1,3,5,10$ (from bottom to top). For the noisier bath (right plot), the characteristic times are very small and the non-Gaussianity is rapidly going to zero in the classicality zone even it is appreciable at the threshold 
$\gamma t_c=0.182$.}
\label{fig-3}
\end{figure*}
The agreement between the Hilbert-Schmidt degree of non-Gaussianity and the other two distance-type measures we discuss here is more accurate during the interaction with noisier reservoirs, as shown in Figs. 2 and 3. All plots showing the time development 
of $\delta_{F}$ and $\delta_{\mathrm{RE}}$ in Figs. 1 and 2 
have a similar aspect which indicates a good consistency of these two measures coming from their monotonicity under the damping map. We expect they induce the same ordering of non-Gaussianity. 
This property is not shared by the Hilbert-Schmidt degree 
of non-Gaussianity and we have to remark the different concavity 
of the plots in the left-hand side of Fig. 3. For weak damping this measure could not induce the same ordering of the non-Gaussianity 
as the relative entropy or Bures distance. In Fig.4 (left) 
we have plotted the entropic measure as a function 
of the fidelity-based degree for the set of damped states 
with $\bar n_R=0.1,\; \bar n=1,$ and different values of $M$. 
The two measures show here a good consistency by preserving 
the same ordering of non-Gaussianity. Also plotted in Fig.4 is 
$\delta_{\mathrm{HS}}$ as a function of $\delta_{\mathrm{RE}}$ 
and $\delta_{F}$ for the same sets of states as in the left panel. 
\begin{figure*}[t]
\center
\includegraphics[width=5cm]{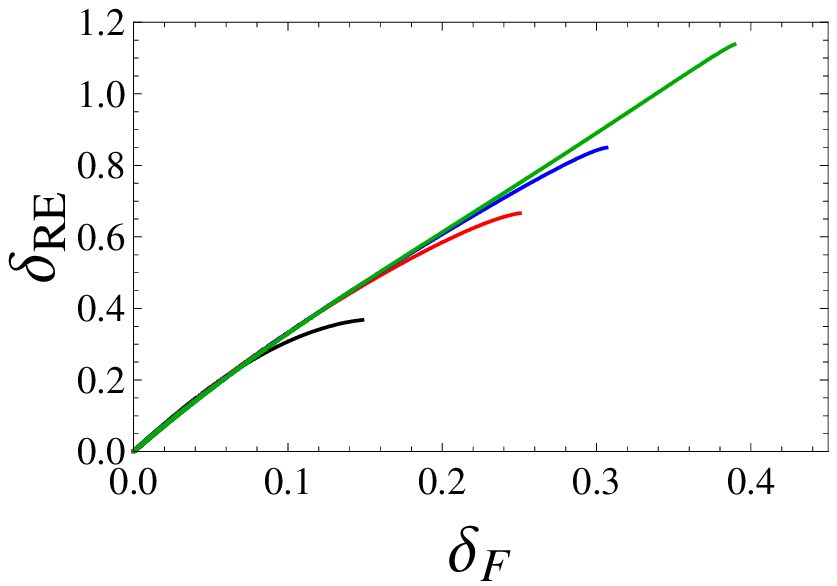}
\includegraphics[width=5cm]{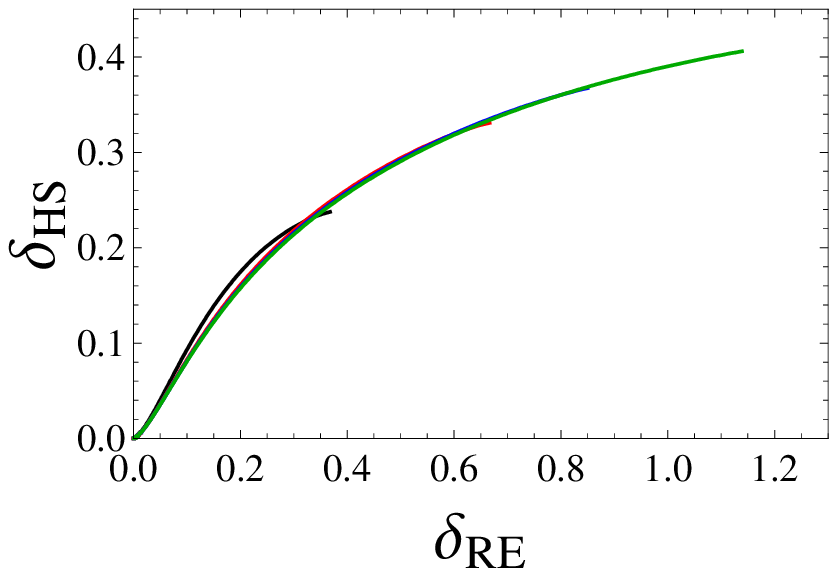}
\includegraphics[width=5cm]{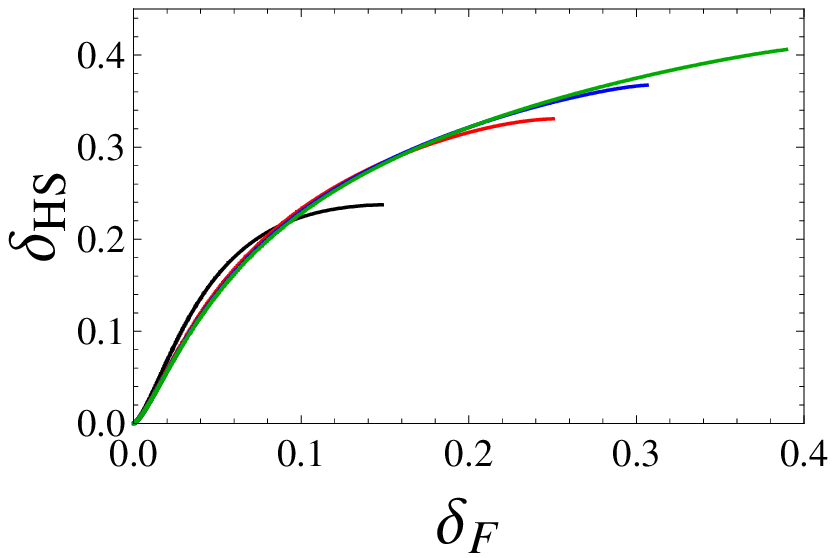}
\caption{Displaying a full consistency between  
$\delta_{\mathrm{RE}}$ and $\delta_{F}$ (left panel) and a partial one between these two measures and the Hilbert-Schmidt one for input $M$-photon added thermal states  ($\bar n=1$) evolving in contact with a thermal bath when  $\bar n_R=0.1$. We used $M=1,3,5,10$ (from bottom to top).}
\label{fig-4}
\end{figure*}
We notice that both of them have the same concavity but they lack consistency for some values of the parameters. Indeed, as shown 
by Figs. 1-3, at a given time the all three measures 
are increasing with the number $M$. This is fairly displayed 
by the left plot in Fig.4 but not entirely by the other two. 
We see that the non-Gaussianity of the states with $M=1$ appears 
to be larger than in the case of $M=10$.

\subsection{Gaussification and entropy production}

The monotonic decay of all non-Gaussianity degrees under damping
does not tell us too much about the loss of nonclassicality 
of the states\ (\ref{PATS}). Decreasing of negativity regions 
of quasiprobability densities seems to be a faithful indicator 
of nonclassicality decay as shown on very general grounds 
in Sec. III. Some time ago, another interesting evolution 
was found for two functions related to the von Neumann entropy, namely, the linear entropy and the 2-entropy \cite{PT00a,PT00b}. 
In Ref.\cite{PT93b} it was shown that a {\em nonclassical} 
Gaussian state under damping displays a maximum in its linear entropy provided that its initial mean occupancy 
$\langle \hat N \rangle |_0$ exceeds the mean occupancy 
of the reservoir $\bar n_R$.  See also a more recent analysis 
in Ref.\cite{D08}. For some pure nonclassical input states, 
a number state in Ref.\cite{PT00a} and an even coherent state in Ref.\cite{PT00b}, it was found analytically that a maximum in the evolution of the $2$-entropy exists under the same mean-occupancy condition.  It appears to us that the existence of a maximum 
in the evolution of the entropy under damping is primarily conditioned by the nonclassicality (expressed in terms of a not well-behaved $P$ representation) of the input state. For instance, the entropy of a classical Gaussian state does not exhibit 
a maximum under damping in any case. But this statement was not proven for mixed non-Gaussian states so far.

The mixed states we are dealing with in this paper are both 
nonclassical and non-Gaussian. We thus find it interesting 
to check on their entropic evolution under damping because 
we can take advantage of having evaluated the von Neumann entropy for both the time-dependent state $\hat \rho^{(M)}(t)$ 
and the reference Gaussian one $\hat \rho^{(M)}_G(t)$.  
In the latter case, the field-reservoir interaction preserves 
the feature of $\hat \rho^{(M)}_G(t)$ being a thermal state. 
Its mean photon number $\langle \hat N \rangle |_t$ 
[Eq.\ (\ref{num1})] varies monotonically from the input value  
$\langle \hat N \rangle |_0=M(\bar n+1)+\bar n $ to the reservoir mean occupancy $\bar n_R$. Therefore, the von Neumann entropy ${\cal S}(\hat \rho^{(M)}_G(t))$ is  a monotonic function of time between the corresponding imposed limits  at $t=0$ and $t\rightarrow \infty$. Good examples of this monotonic evolution 
are shown on the left-hand side of Fig.5 for a noisy reservoir 
with $\bar n_R=5$. 
\begin{figure*}[h,t,b]
\center
\includegraphics[width=7cm]{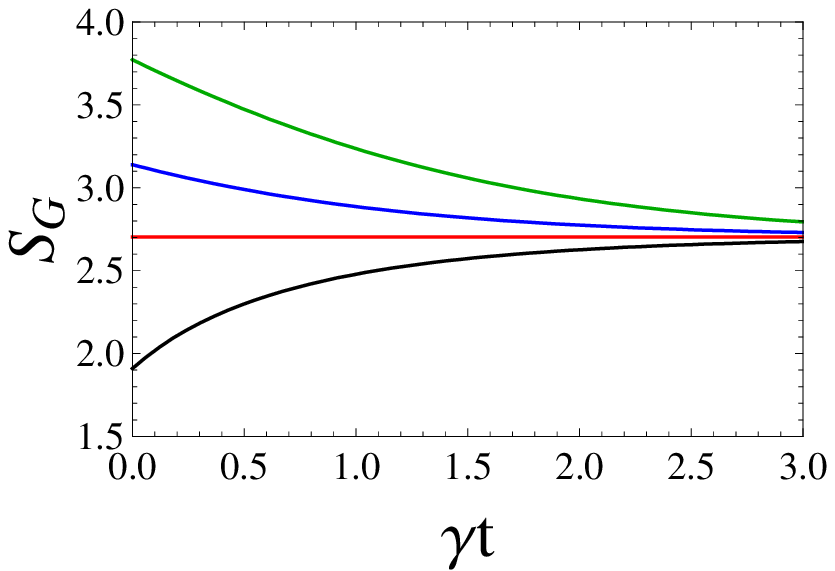}
\includegraphics[width=7cm]{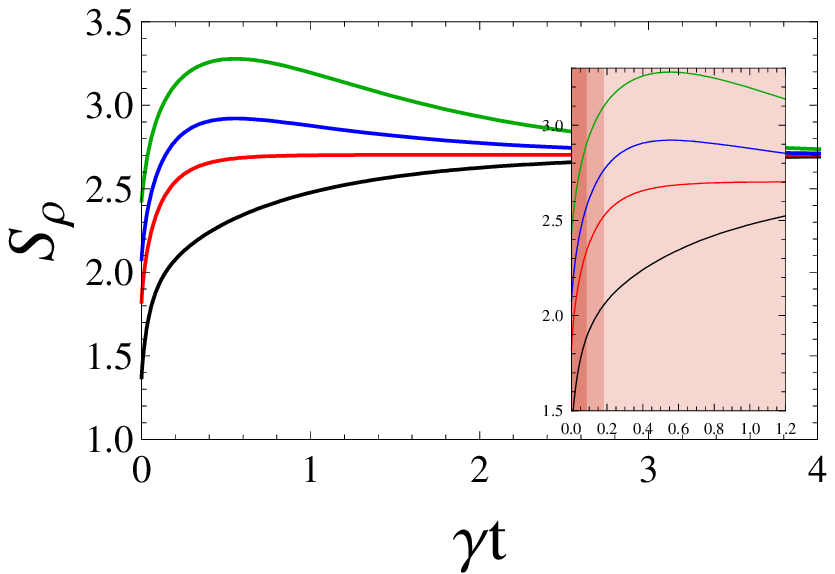}
\caption{Entropy production during the Gaussification presented 
in Fig. 2, namely: $\bar n_R=5,\; \bar n=0.5$, and $M=1,3,5,10$ 
(from bottom to top). The inset in the right plot presents also 
the three regions of decoherence defined by the characteristic parameters of the bath $\gamma t_w=0.087$ and $\gamma t_c=0.182 $. The maximum of the entropy is reached in the classicality region. See Table I.}
\label{fig-5}
\end{figure*}
The input mean occupancy $\langle \hat N \rangle |_0$ 
for the states with $\bar n=0.5$ and various $M$ are placed in Table I together with the corresponding values of the entropy 
for the reference Gaussian states $\hat \rho^{(M)}_G(0)$ and the given ones $\hat \rho^{(M)}(0)$. To calculate the latter ones, 
we made use of the density matrix\ (\ref{dpa}). Notice that in 
all the chosen cases the entropy of the input $\hat \rho^{(M)}(0)$ 
is smaller than that of the reservoir. The von Neumann entropy 
of the damped state $\hat \rho^{(M)}(t)$ is plotted on the 
right-hand side of Fig. 5 for the same parameters. 

\begin{table}[h]
\center
\begin{tabular}{|c|c|c|c|c|}
\hline
M&$\langle \hat{N} \rangle \vert_0 $&${\cal S}(\hat \rho^{(M)}_G)$
&${\cal S}(\hat \rho^{(M)})$&$\gamma t_{max}$\\
\hline
1& 2&1.909&1.372&-\\
\hline
3& 5&2.703&1.824&-\\
\hline
5& 8&3.140&2.078&0.5562\\
\hline
10&15.5&3.770&2.429&0.5538\\
\hline
\end{tabular}
\caption{Entropic characterization of the input states whose Gaussification is analyzed in Figs.2 and 3, namely having $\bar n=0.5.$ The thermal mean occupancy of the reservoir is 
$\bar n_R=5.$ In the last column we have inserted the position 
of the maximum of the entropy.}
\label{tab1e}
\end{table}
By contrast to its monotonic evolution for 
$\langle \hat N \rangle |_0 \leqq \bar n_R$ (cases $M=1, 3$), 
we point out the existence of a short-time maximum 
of the von Neumann entropy ${\cal S}(\hat \rho^{(M)}(t))$ when 
$\langle \hat N \rangle |_0 >\bar n_R$ (cases $M=5,10$).
The maximum is reached at a time depending slightly on $M$ and situated quite deeply in the classicality region. Therefore, 
the evolution of the von Neumann entropy follows the same pattern as in the case of mixed Gaussian states and obeys 
the same mean-occupancy and nonclassicality conditions.

\section{Discussion and conclusions}

To sum up, in this work we have investigated the evolution of 
non-Gaussian one-mode states during the interaction of the field mode with a thermal reservoir resulting in two processes:  
Loss of nonclassicality indicated by the time development of the Wigner and $P$ functions and loss of non-Gaussianity shown 
by some recently introduced distance-type measures. We have analytically proved that the evolution of all input Fock-diagonal states is characterized by three decoherence regions that can be related to some recent findings in quantum computation 
\cite{VWFE}.  The end of the first one at the time $t_w$ 
[Eq.\ (\ref{tw})] represents the onset of the positivity of the Wigner function. The damped states belonging to the first region are nonclassical and can be used for quantum computational speed-up. The second region ends at the time $t_c$ [Eq.\ (\ref{tc})]  when the $P$ representation becomes positive. The states in this region were described in quantum computation as bound universal states. The third one is a classicality domain towards the limit 
$t\rightarrow \infty$ when the mode shares the thermal state 
of the reservoir. States damped after the moment $t_c$ can be efficiently simulated on a classical computer because they possess a positive Wigner function \cite{ME, VFGE}.  The $M$-photon-added thermal states perfectly supported these general findings on the nonclassicality decay. For these specific states we have also found that decaying of non-Gaussianity was consistently described 
by the three distance-type measures used in this paper. Even 
with some alterations of ordering, the fidelity-based degree, 
the Hilbert-Schmidt one, and the entropic measure evolve monotonically, as expected for any good measure of non-Gaussianity \cite{GMM}. We have also marked on all time-development plots 
in Figs.1-3 the three decoherence zones discussed in Sec. IV A, which are delimitated by the thresholds $\gamma t_w$
[Eq.\ (\ref{tw})] for the onset of the positivity of the Wigner function and $\gamma t_c$ [Eq.\ (\ref{tc})] representing 
the onset of classicality.  In the case of weak damping 
(small $\bar n_R$),  the non-Gaussianity is appreciable
for $t<t_c$ and slowly decays for  $t>t_c$.  For noisier baths, 
the characteristic times are very small and the non-Gaussianity 
is rapidly going to zero in the classicality zone.

As a side effect, we pointed out a nonmonotonic time development 
of the von Neumann entropy whose behavior depends on the relation between the input mean occupancy $\langle \hat N \rangle |_0$ 
of the state and the thermal mean occupancy $\bar n_R$ 
of the reservoir. States with input mean occupancies greater 
than $\bar n_R$, but with input entropy not exceeding that 
of the reservoir, were found to present a transient mixing enhancement, visible as a maximum in the evolution of their entropy. 

\section*{ACKNOWLEDGMENT}

This work was supported  by the Romanian National Authority for Scientific Research through Grant No.~PN-II-ID-PCE-2011-3-1012 
for the University of Bucharest.

\appendix*
\section{Some power series involving hypergeometric functions}
A Gauss hypergeometric function is the sum of the corresponding hypergeometric series,
\begin{equation}
_{2}F_{1}(a, b; c ; z):=\sum\limits_{n=0}^{\infty}\frac{(a)_{n}
(b)_{n}}{(c)_{n}}\frac{z^n}{n!}\,\qquad (|z|<1), 
\label{a1}
\end{equation}
where $(a)_{n}:=\Gamma(a+n)/\Gamma(a)$ is Pochhammer's symbol
standing for a rising factorial. This definition is extended by analytic continuation. The confluent (Kummer) hypergeometric 
function has the Maclaurin series:  
\begin{equation}
_{1}F_{1}(a; c; z):=\sum\limits_{n=0}^{\infty}\frac{(a)_{n}}
{(c)_{n}}\frac{z^n}{n!}.
\label{a11}
\end{equation}
In Sec. III, we have used Humbert's summation formula \cite{24}
\begin{align}
\exp(-xz)\, _{1}F_{1}(a; c; z)=\sum_{p=0}^{\infty}\frac{(-xz)^{p}}{p!} \,_{2}F_{1}\left( -p, a; c; \frac{1}{x}\right) .
\label{a55}
\end{align}
Note that a Laguerre polynomial of degree $M$ is a confluent hypergeometric  function, 
\begin{equation}
L_M(z)= {_{1}F_{1}} (-M;\, 1;\, z),
\label{lag}
\end{equation}
while a Legendre polynomial of degree $M$ can be expressed 
in terms of a Gauss hypergeometric function:
\begin{align}
P_M(z)=\left(\frac{z+1}{2}\right)^M\; & {_{2}F_{1}}
\left( -M, -M;\, 1; \, \frac{z-1}{z+1} \right)  
\notag \\ 
& (M=0,1,2,3, \ldots ). 
\label{a2}
\end{align}
Let us consider a well-known generating function of a class 
of hypergeometric polynomials \cite{HTF251}:
\begin{align} 
G_0(u,z) & :=\sum_{l=0}^{\infty}\, u^{l}\;_{2}F_{1}(-l, a; 1; z)
\notag \\
& =(1-u)^{a-1}(1-u+u z)^{-a} 
\notag \\
& (|u|<1, \quad |u(1-z)|<1).
\label{a3}
\end{align}
In the body of the paper we need to evaluate the sum
\begin{align}
G_n(u,z): & =\sum_{l=n}^{\infty}\, \binom{l}{n}\, u^l\;
_{2}F_{1}(-l, a; 1; z) \notag \\
& =\frac{1}{n!}\; u^n \;\frac{\partial^n G_0(u,z)}{\partial u^n}.
\label{a4}
\end{align}
We have performed the sum \eqref{a4} in two ways. First, we have applied an elegant method described in the recent paper \cite{pol} to find the $n$th-order derivative of such a generating function using Cauchy's integral formula. Second, because the right-hand side of Eq.\ (\ref{a3}) has a simple structure, the $n$th-order derivative can be put in a closed form after some routine algebra. With both methods we got the formula
\begin{align}
G_n(u,z) & =u^n (1-u)^{a-n-1}(1-u+u z)^{-a}
\notag \\
& \times{_{2}F_{1}}\left(-n, a; 1; \frac{z}{1-u+u z}\right).
\label{a5}
\end{align}
In evaluating the quasiprobability distributions $W(\beta;s)$, 
we employed the sum of a nontrivial expansion in terms of Laguerre polynomials whose derivation can be found in Ref.\cite{pol}:
\begin{align}
\sum_{l=n}^{\infty}\,\binom{l}{n}\, u^l\, L_l(z) & =
\frac{u^n}{(1-u)^{n+1}}\, \exp{\left(-\frac{u z}{1-u}\right)}
\notag \\
& \times L_{n}\left( \frac{z}{1-u} \right), \quad  (|u|<1).
\label{la5}
\end{align}
The known summation of a power series whose coefficients are proportional to the product of two Gauss hypergeometric 
polynomials \cite{HTF252} was also employed in Sec. IV:
\begin{align}
& \sum_{n=0}^{\infty}\, \frac{(c)_{n}}{n!}(- v)^{n}
\,_{2}F_{1}(-n, b; c; z)\,_{2}F_{1}(-n, b^{\prime}; c; z^{\prime})
\notag \\  
& =(1+v)^{-c+b+b^{\prime}}
(1+v-vz)^{-b}\,(1+v-vz^{\prime})^{-b^{\prime}} 
\notag \\ 
& \times  _{2}F_{1}\left[ b,\, b^{\prime};\, c;\,
\frac{-v zz^{\prime}}{(1+v-vz)(1+v-vz^{\prime})}\right] .
\label{a7}
\end{align}
In our case, the parameters take particular values,
$$b^{\prime}=b,\,c=1,\,z^{\prime}=z,$$ 
so that Eq.\ (\ref{a7}) reduces to a slightly simpler one:
\begin{align}
& \sum_{n=0}^{\infty}v^n \,\left[ _{2}F_{1}(-n, b;\, 1;\, z)
\right] ^2=(1-v)^{-1+2 b}(1-v+v z)^{-2 b}
\notag \\ 
& \times {_{2}F_{1}}\left[ b, b;\, 1;\, \frac{v z^2}
{(1-v+vz)^2}\right] .
\label{a77}
\end{align}

%\section*{References}

\end{document}